\documentclass[twocolumn,final,aps,prl,showpacs,superscriptaddress,amsmath,amssymb,amsfonts,floatfix]{revtex4-1}

\usepackage{graphicx}
\usepackage{multirow}
\usepackage{epsfig}
\usepackage{url}
\usepackage{color}
\usepackage{grffile}

\usepackage[colorlinks,breaklinks,bookmarks=true,citecolor=blue,linkcolor=blue,urlcolor=blue]{hyperref}
\usepackage{color}
\usepackage{tabularx}
\usepackage{sidecap}
\usepackage{soul}
\usepackage{float}

\usepackage{orcidlink}

\begin{document}

\title{Effects of different concentrations of topotactic hydrogen impurities on the electronic structure of nickelate superconductors}

\author{Chenye Qin}
\affiliation{Institute of Theoretical and Applied Physics, Jiangsu Key Laboratory of Thin Films, School of Physical Science and Technology, Soochow University, Suzhou 215006, China}

\author{Mi Jiang\,\orcidlink{0000-0002-9500-202X}}
\email{jiangmi@suda.edu.cn}
\affiliation{Institute of Theoretical and Applied Physics, Jiangsu Key Laboratory of Thin Films, School of Physical Science and Technology, Soochow University, Suzhou 215006, China}

\author{Liang Si\,\orcidlink{0000-0003-4709-6882}}
\email{liang.si@ifp.tuwien.ac.at}
\affiliation{School of Physics, Northwest University, Xi'an 710127, China}
\affiliation{Institute of Solid State Physics, TU Wien, 1040 Vienna, Austria}


\begin{abstract}
Infinite-layer nickelate superconductors have recently been discovered to share both similarities and differences with cuprate superconductors. Notably, the incorporation of hydrogen (H) through topotactic reduction has been found to play a critical role in their electronic structure and, consequently, their superconductivity. In this study, we utilized a theoretical approach combining density-functional theory and impurity approximation to design three characteristic multi-orbital Hubbard models representing low, moderate, and high concentrations of topotactic-hydrogen. Consistent with experimental findings, our simulations revealed that both low and high concentrations of topotactic-hydrogen induce high-spin states ($S$=1) that are composed by holes at $d_{x^2-y^2}$ and $d_{z^2}$ orbitals and consequently the emergent inter-site hopping between $d_{z^2}$ to $d_{x^2-y^2}$ is unfavorable for superconductivity. Conversely, an optimal concentration of 25\% H aligns with the single Ni-$d_{x^2-y^2}$ band picture of superconductivity in infinite-layer nickelates, demonstrating its beneficial effect on promoting superconducting behavior.
\end{abstract}

\maketitle

\section{I.~Introduction}

The pairing mechanism driving the unconventional superconductivity (SC) remains highly controversial several decades after the initial discovery of cuprate superconductors \cite{Bednorz1986}. Different proposals have been proposed, such as antiferromagnetic (AFM) spin \cite{Monthoux1991,Scalapino12,Vilardi2020,Sordi2012,Gull2015,Kitatani2019}, charge \cite{Castellani1997}, and current fluctuations \cite{Varma1997}.
Early theoretical attempts had predicted the possibility of nickelate superconductors \cite{Anisimov1999} and heterostructures \cite{Chaloupka2008,PhysRevLett.103.016401,Hansmann2010b}.
In combination with oxygen, Ni favors an octahedral configuration with a Ni$^{2+}$ or Ni$^{3+}$ oxidation state, although a planar coordination and Ni$^{1+}$ oxidation state can also be found in the infinite-layer $R$NiO$_2$ system.

By employing the topotactic chemical reduction method, infinite-layer nickelate superconductors have been discovered recently \cite{li2019superconductivity,Osada2020,Zeng2021,Osada2021}, which have a similar crystal structure to cuprates but with a different set of electronic properties \cite{li2019superconductivity,zeng2020,Osada2020,Zeng2021,Osada2021,pan2021,PhysRevLett.125.077003,Nomura2019,Jiang2019,Werner2019,Lechermann2019,Lechermann2020,Petocchi2020,Adhikary2020,Motoaki2019,hu2019twoband,Wu2019,Karp2020,Kitatani2020,Zhang2019,Subhadeep2020,jiang2019electronic,Si2019,Geisler2021,Klett2022,LaBolitta2022,ji2021superconductivity,Higashi2021,lee2022character} on both infinite- and finite-layer nickelates \cite{pan2021,Worm2021c,PhysRevB.106.115132}.
For the nature of the nickelate SC, whether it is driven by multi-band \cite{PhysRevB.102.161118,PhysRevLett.125.077003,PhysRevB.104.144504,Nomura2019,Jiang2019,Werner2019,Lechermann2019,Lechermann2020,PhysRevMaterials.5.044803,PhysRevLett.129.077002,Petocchi2020,Adhikary2020,Gu2020} or single-band \cite{Wu2019,PhysRevB.104.L220505,Held2022,harvey2022evidence,Wu2019,Karp2020,Kitatani2020,Worm2021c,kitatani2022optimizing,kitatani2023ab}, $d$-wave or $s$-wave pairing symmetry\cite{Gu2020b,PhysRevB.102.220501,harvey2022evidence,rossi2022broken,chow2022pairing}, Kondo effect and the role of $f$-orbital \cite{Zhang2019,gu2020substantial,zhang2021magnetic,Subhadeep2020,PhysRevB.104.165137}, charge order \cite{tam2021charge,krieger2021charge,rossi2022broken,chen2022charge}, magnetic ground-state  \cite{Cui2021,fowlie2022intrinsic,PhysRevResearch.4.023093,Lu2021,lin2022universal}, and whether topotactic hydrogen (H) is destructive or essential \cite{Si2019,Malyi2022,Puphal2022,PhysRevMaterials.6.044807,Cui2021,Puphal2022,ding2023critical} to SC, are still on debate.
Furthermore, the presence of possible atomic \cite{si2022fingerprints} and structural defects \cite{lee2022character}, such as additionally oxygen or shear fault, can greatly influence the electronic and structural properties of infinite-layer nickelate superconductors.

Theoretical simulations~\cite{Si2019,Malyi2022,PhysRevMaterials.6.044807,ding2023critical} have shed light on the fact that the topotactic hydrogen modifies the doping degree and low-energy band topology near Fermi surface (FS), and alters the electronic and SC properties of infinite-layer nickelates \cite{Si2019}.
Recently, Ding and coworkers \cite{ding2023critical} reported the measurement of a strong dependence of SC in Nd$_{0.8}$Sr$_{0.2}$NiO$_2$H$_{\delta}$ on the concentration of topotactic-H $\delta$. In particular, the $T_c$ of $\sim$10\,K is observed in a narrow range of H concentration $\delta\sim$0.22-0.27, peaking around 0.25, leading to the SC dome versus both Sr doping and concentration of H.

Although the presence of topotactic-H is experimentally confirmed, its alleged importance or even necessity for SC has never been discussed carefully in the literature. In fact, while the mechanism behind SC in infinite-layer nickelates remains debated, none of the proposed ones require the presence of H \cite{Kitatani2020,PhysRevLett.129.077002,li2022two,chow2022pairing,PhysRevB.102.220501}.
On the contrary, if one adopts the perspective of nickelates as cuprate analogs, the doped hydrogen, tending to form H$^-$ so that locally induces a Ni$^{2+}$ (3$d^8$) configuration, is expected to be detrimental to SC. 
This hence raises the opening question of whether the topotactic-H is able to  destroy the observed SC or even induce another superconductive resource such the electron-phonon coupling that may be responsible for the observed unusual nodeless $s$-wave gap \cite{chow2022pairing,Gu2020b}.

\begin{figure}[t!]
\psfig{figure=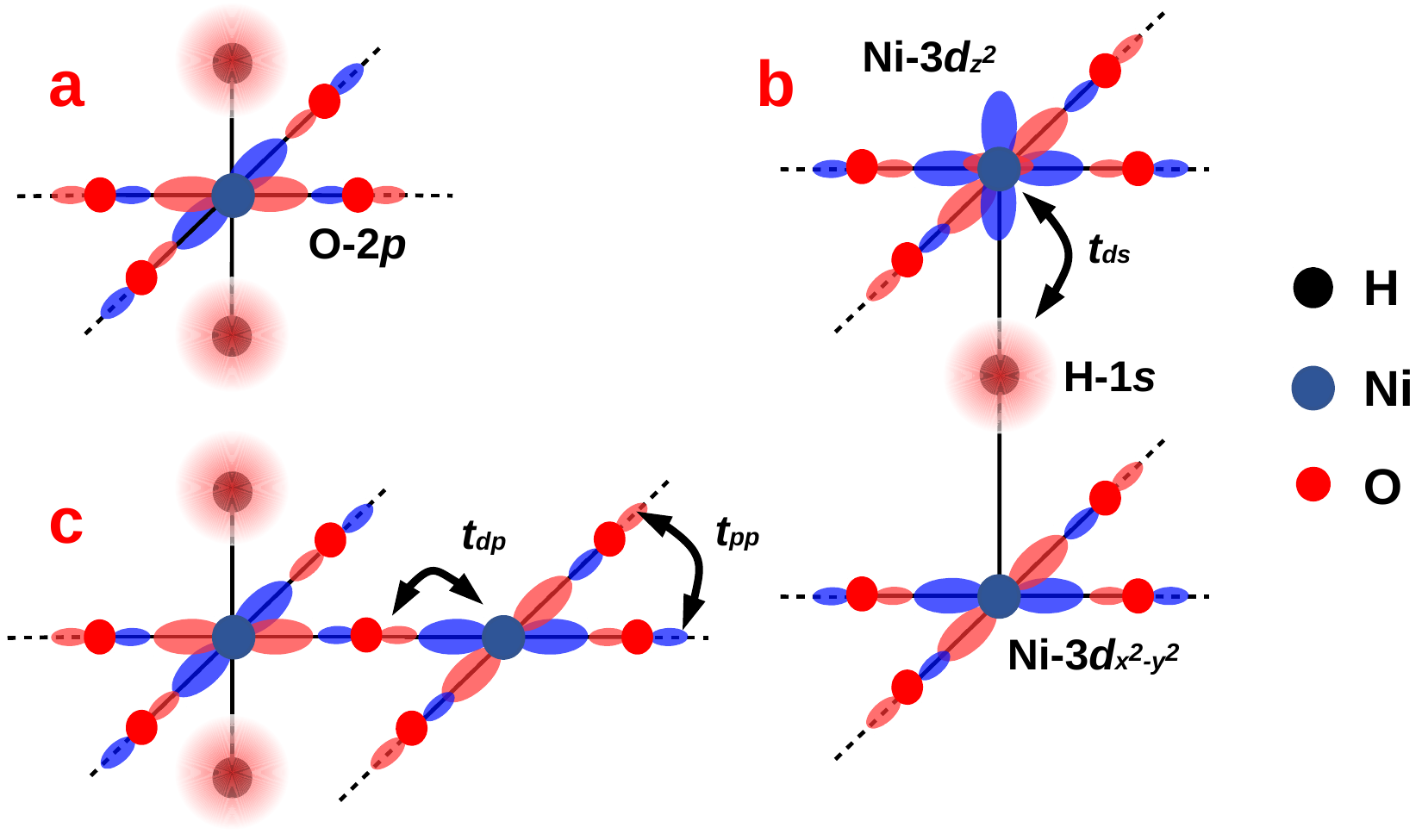,width=.45\textwidth, clip}
\caption{Schematic geometry of employed models : (a) single Ni site with two topotactic H sites that represents 100\% doped H case; (b) slight H-doped case with a single topotactic H that is surrounded by two Ni sites; (c) coexistence of Ni with and without topotactic H that represents 50\% moderately doped H case. Note that all five $3d$ orbitals are incorporated for Ni impurity in all models  despite that here we only illustrate $d_{z^2}$ orbital for one impurity.}
\label{geom}
\end{figure}

To investigate these questions, we explored the electronic structure of a multi-orbital Hubbard model using density-functional calculations (DFT) \cite{PhysRev.136.B864} in combination with the impurity approximation. In this approach, we treated Ni and H as impurities embedded in the O lattice, as depicted in Fig.~\ref{geom}(a-c) [the panels (a), (b) and (c) represent the system with high (100\%), low and middle concentrations of topotactic H, respectively], to examine the influence of topotactic H doping concentration on SC. 
Our findings indicate that the intecalation of H plays important role in hole distribution between orbitals and thereby electronic structure. The parameters of the newly emergent physical degree of freedom, including the hopping between Ni-$d_{z^2}$ and H-1$s$ ($t_{ds}$), and the enhancement of the on-site energy of Ni-$d_{z2}$ are identified as the most important two parameters determining the phase diagram of our models with various concentrations of topotactic H. Within the DFT computed parameters region, both high [Fig.~\ref{geom}(a)] and low [Fig.~\ref{geom}(b)] concentrations of H are detrimental to SC due to the induced high-spin triplet states that are characterized by the holes distributed at not only $d_{x^2-y^2}$ but also $d_{z^2}$.
On the contrary, at the optimal concentration (approximately 25\%), an in-plane ordered formation and distribution of one-dimensional H-chains perpendicular to the Ni-O plane protect the $S$=1/2 state with holes only residing on the $d_{x^2-y^2}$ orbital [Fig.~\ref{geom}(c)]. This arrangement effectively limits the impact of topotactic-H to Ni sites situated nearest to H atoms, facilitating long-range effective inter-site hopping of holes along both the $x$ and $y$ directions, thus promoting SC. Finally, higher concentration, e.g.~50\%, is expected to eliminate the effective inter-site electron hopping between $S$=1/2 $d_{x^2-y^2}$ by the emergence of inter-site hopping between $d_{x^2-y^2}$ to $d_{z^2}$.

\section{II.~Model and Method}

In this study, we build upon our previous research~\cite{Mi2020,Mi2020a,Mi2022}, which focused on hole doping in a system consisting of a single Ni$^{1+}$ ($3d^9$) impurity properly embedded in an infinite square lattice of O $2p^6$ ions. To investigate the effects of H doping, we designed three distinct models representing situations of 100\%, moderate, and tiny H dopings, as illustrated in Figure~\ref{geom}(a-c).
For 100\% H doping, corresponding to a fully topotactic H compound, we believe that H-chains~\cite{PhysRevB.107.165116} form perpendicular to the NiO$_2$ plane, as depicted in Fig.~\ref{geom}(a).
Fig.~\ref{geom}(b) shows the model for tiny H doping achieved by inserting a single H into a bilayer Ni-O system.
Finally, Fig.~\ref{geom}(c) presents the situation of moderate ($\sim$ 50\%) H doping, where the coexistence of normal and H-doped regions is simulated by incorporating two distinct Ni impurities: one hybridized with the H chain and the other not.

The vacuum state is defined as $3d^9 2p^6 1s^2$, representing a single hole on the Ni-$3d$ orbital and fully occupied O-$2p$ and H-$1s$ orbitals. In other words, the topotactic H is initially treated as H$^-$ in all three models [Fig.~\ref{geom}(a-c)]. Therefore, without any additional hole doping, the undoped states correspond to having (a) two, (b) three, and (c) three holes, respectively, due to the presence of two impurities in the latter two models. It is important to note that in the undoped systems, the self-doping effect of topotactic H onto the Ni-O planes has already been taken into account by considering the hopping between H-1$s$ and Ni-$d_{z^2}$: $t_{ds}$ and between H-1$s$ and O-$p$: $t_{ps}$. In addition to the undoped systems, this work will also investigate the scenarios with one additional hole doped into each system, resulting in an increase in the number of holes by one for each case.

The general Hamiltonian reads as:
\begin{equation}\label{Ham} {\cal H} =
\hat{U}_{dd}+\hat{T}_{pd}+\hat{T}_{pp}+\hat{T}_{ds}+\hat{T}_{ps}+\hat{T}_{z}+\hat{E}
\end{equation}
where $\hat{U}_{dd}$ includes all Coulomb and exchange interaction of the $3d^8$ multiplet corresponding to $D_{4h}$ symmetry in terms of Racah parameters $A,B,C$, which are linear combinations of conventional Slater integrals.
$ \hat{T}_{pd}$, $\hat{T}_{pp}$, $\hat{T}_{ds}$, $\hat{T}_{ps}$ incorporate hopping integrals between the Ni-$3d$ orbitals, adjacent O-$2p$ ligand orbitals ($L$ denotes the four ligand O orbitals nearest to the impurity), and doped H-$s$ orbital ($H$ the hole located at H-1$s$). In addition, for model Fig.~\ref{geom}(b) with bilayer NiO$_2$, there is still a renormalized direct interlayer hybridization between Ni-3$d_{z^2}$ orbitals incorporated in $\hat{T}_{z}$ despite of its much smaller magnitude than $\hat{T}_{ds}$.
In other words, this model explicitly considers the effect of interstitial H, while in our previous work~\cite{Mi2023}, the impact of H was accounted for in an implicit manner.
Besides, $\hat{E}$ describes the site energies of various Ni-$3d$ and O-$2p$, H-$s$ orbitals. 

These designed models are reminiscent of our previous studied single and bilayer systems relevant for infinite-layer nickelates~\cite{Mi2020,Mi2020a,Mi2022,Mi2023}. 
The analysis is achieved by performing exact diagonalization (ED) to determine the nature of the GS, precisely the weights of different hole configurations in GS, of these distinct models to explore the influence of H doping range. 
To label the multiple-hole states, we use the notation such as $d^8$-$d^9$ or detailed $d_{x^2-y^2}L$-$d_{z^2}$, $d_{z^2}H$-$d_{x^2-y^2}$ etc. to denote the configuration associated with distinct Ni impurities on each layer separated by hyphen.

\begin{table}[h]\label{table}
\footnotesize
\caption{On-site energies $\epsilon$, Racah parameters $A,B,C$, and hopping integrals $T^{pd}_{mn}$ with $m \in \{d_{x^2-y^{2}}, d_{z^2}\}$, $n \in \{p_x, p_y \}$, where $m,n$ are nearest neighbors, extracted from DFT calculations. Note that we only consider $p_x$ and $p_y$ orbitals with lobes pointing to the impurity. For model Fig.~\ref{geom}(c) with bilayer NiO$_2$, there is still direct interlayer hybridization $T^z_{m}$ between Ni-3$d$ orbitals. Only the magnitudes are shown and the sign convention follows the lobes of different orbitals. Note that the DFT values of $t_{ds}$ and $\epsilon(d_{z^2})$ are only for reference and they will be varied as two major control parameters throughout the work. All values are of unit of eV.}

\centering
\begin{tabular}{c|c|c c c c |c} 
 \hline\hline

  $T^{pd}_{x^2-y^{2},n}$  & $T^{pd}_{z^2,n}$  &
  $t_{pd}$ & $t_{pp}$ &  $t_{ds}$ &  $t_{ps}$ &  
  $T^z_{z^2}$  \\ [0.5ex] 
 \hline
 $t_{pd}$ &  $t_{pd}/\sqrt{3}$ & 1.3 &  0.5 & 1.63 & 0.58 & 0.068 \\ 
 \hline 
\end{tabular} 

\centering
\begin{tabular}{c c c c c c | c c c } 
 \hline

  $\epsilon(d_{x^2-y^{2}})$  & 
  $\epsilon(d_{z^2})$  & $\epsilon(d_{xy})$  &
  $\epsilon(d_{xz/yz})$  & $\epsilon_p$ &  $\epsilon_H$ 
  & A  & B  & C \\ [0.5ex] 
 \hline
 0  & 0.35 & 1.55 & 1.9 & 4.7  & 4.8 & 6.0 &  0.15&  0.58  \\
 \hline\hline
\end{tabular} 
\label{table}
\end{table}

All the parameters are listed in Table~\ref{table}. 
The on-site energies and hopping integrals between different orbitals are extracted from the DFT calculation, where the site energy of Ni-$3d_{x^2-y^2}$ is set to be zero as reference.
Besides, the conventional Racah parameters $A=6.0$ eV, $B=0.15$ eV, $C=0.58$ eV describing the Coulomb and exchange interactions are adopted, which is motivated by the assumption that the infinite-layer cuprates and nickelates have similar interaction strength~\cite{Mi2020}.  
The on-site energies and hopping integrals between different orbitals are determined from the DFT \cite{PhysRev.136.B864} calculation and Wannier \cite{PhysRev.52.191,mostofi2008wannier90} projections using \textsc{WIEN2k} \cite{blaha2001wien2k,Schwarz2002} and \textsc{WIEN2WANNIER} \cite{kunevs2010wien2wannier} and revised Perdew-Burke-Ernzerhof for solids (PBESol) of the generalized gradient approximation (GGA) \cite{PhysRevLett.100.136406} for the treatment of exchange-correlations functional. 
The most crucial parameters in all three models are the hybridization $t_{ds}$ between Ni-3$d_{z^2}$ and apical H-1$s$ orbitals. In this study, we will use $t_{ds}$ as a control parameter to simulate the experimental pressure effects, even though Table~\ref{table} provides the realistic value under ambient pressure conditions. 

\begin{figure*}[t]
\psfig{figure=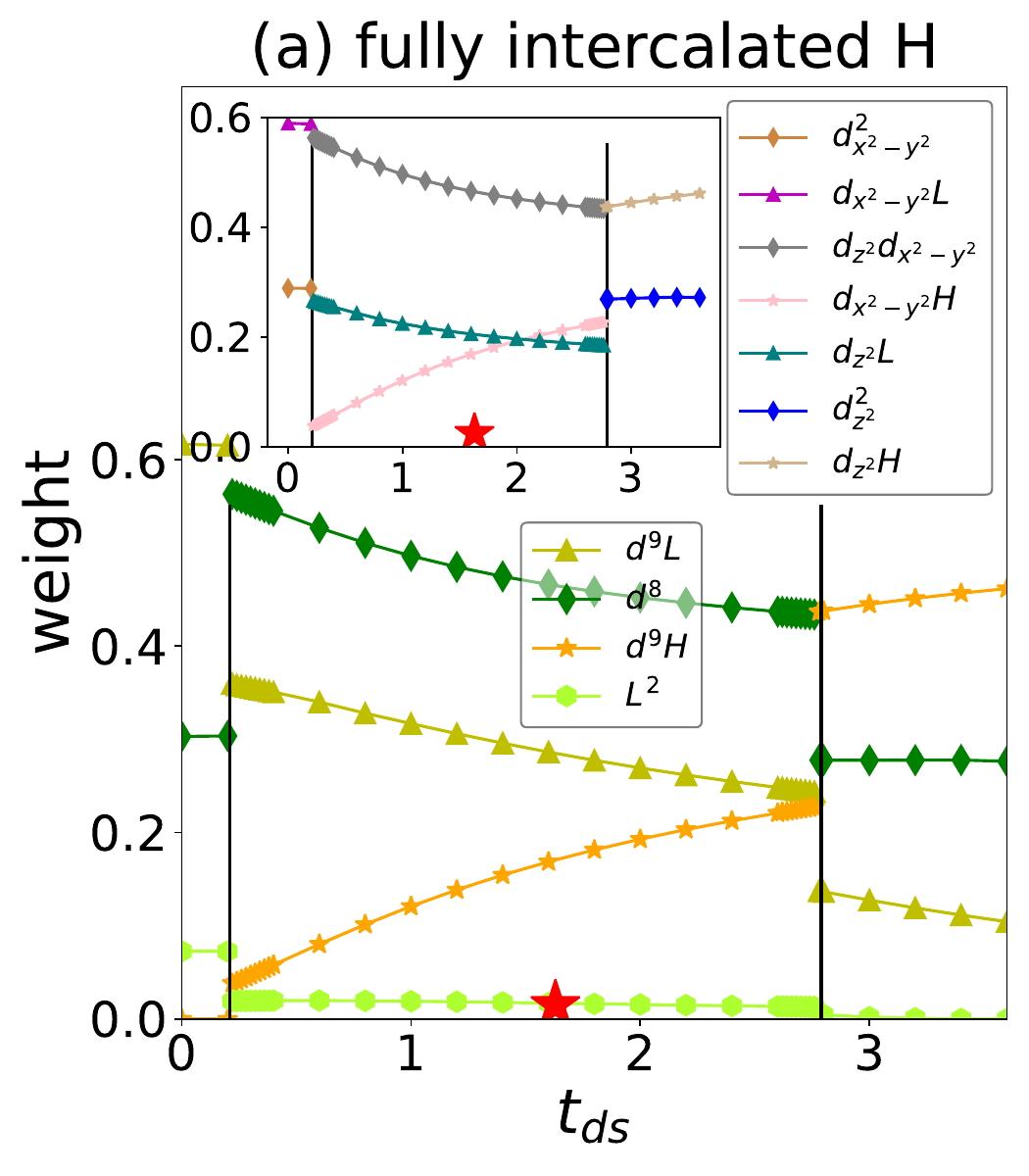,height=6cm,width=.31\textwidth, clip}
\psfig{figure=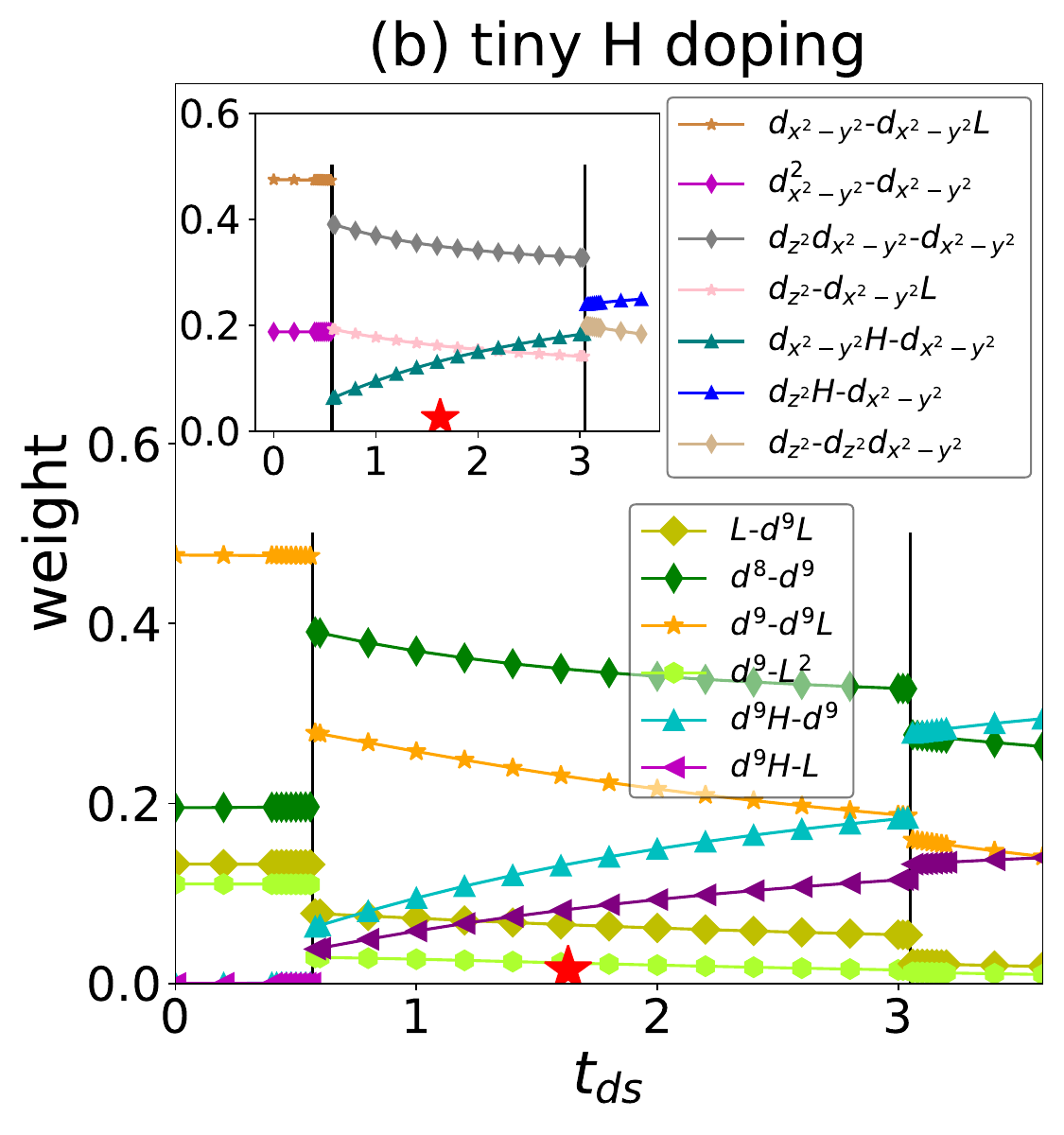,height=6cm,width=.31\textwidth, clip} 
\psfig{figure=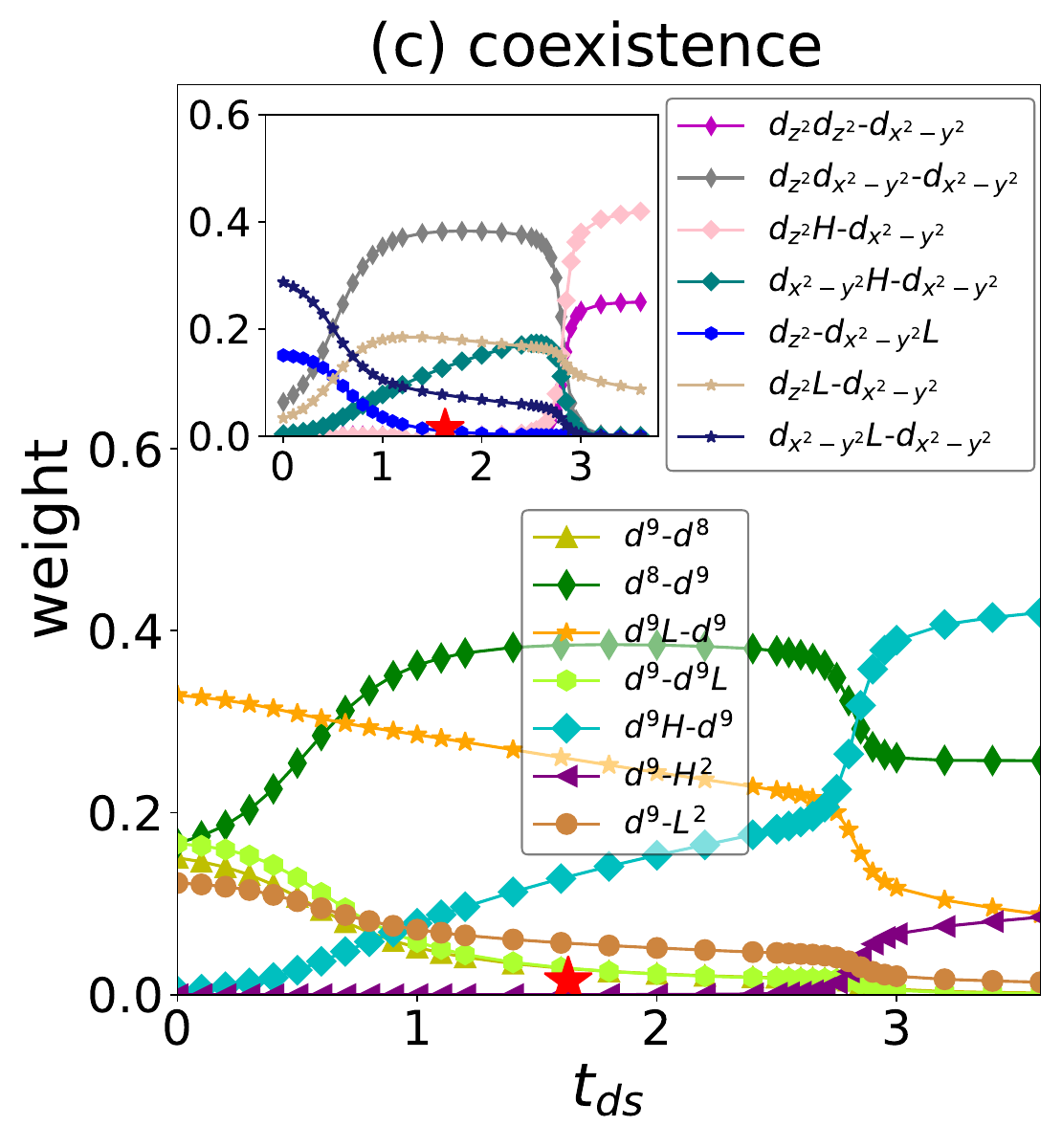,height=6cm,width=.31\textwidth, clip} 

\caption{Evolution of the GS weight distribution in undoped H intercalated systems with increasing the hybridization $t_{ds}$ between Ni-$d_{z^2}$ and H-$s$ orbital for (a) 100\% intercalated H; (b) bilayer NiO$_2$ sandwiching one doped H; (c) coexistence of H doped and normal Ni at fixed $\epsilon_{d_{z^2}}=0.35$ eV. The insets show the decomposed weights of various configurations. The critical $t_{ds}$ is marked by solid black vertical line and the red star denotes the location corresponding to realistic DFT parameter.}
\label{weight}
\end{figure*}

\section{III.~Results and discussion}

\begin{figure*}[t!]
\psfig{figure=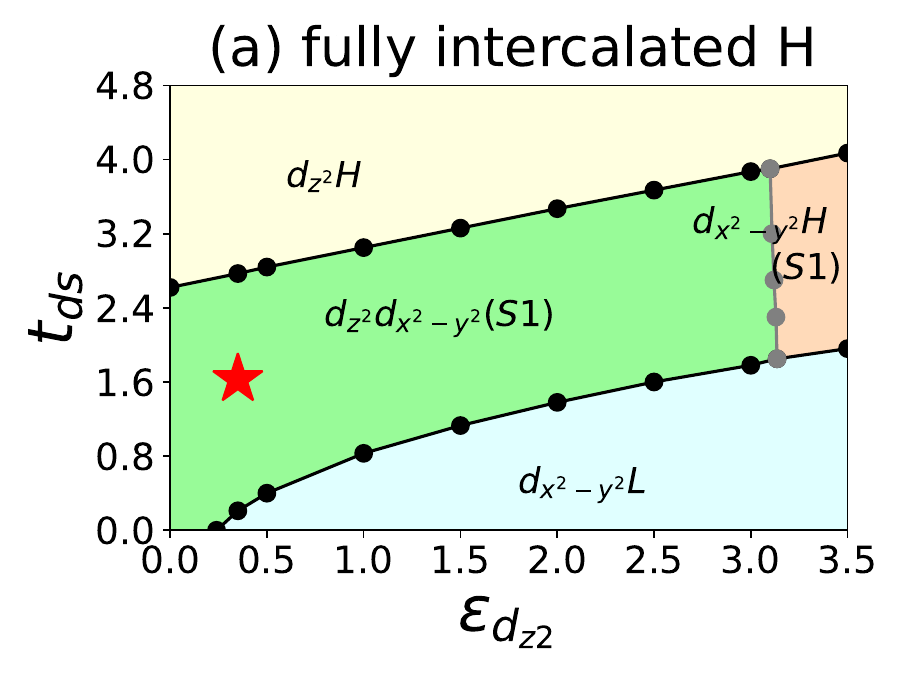,height=4cm,width=.32\textwidth, clip}
\psfig{figure=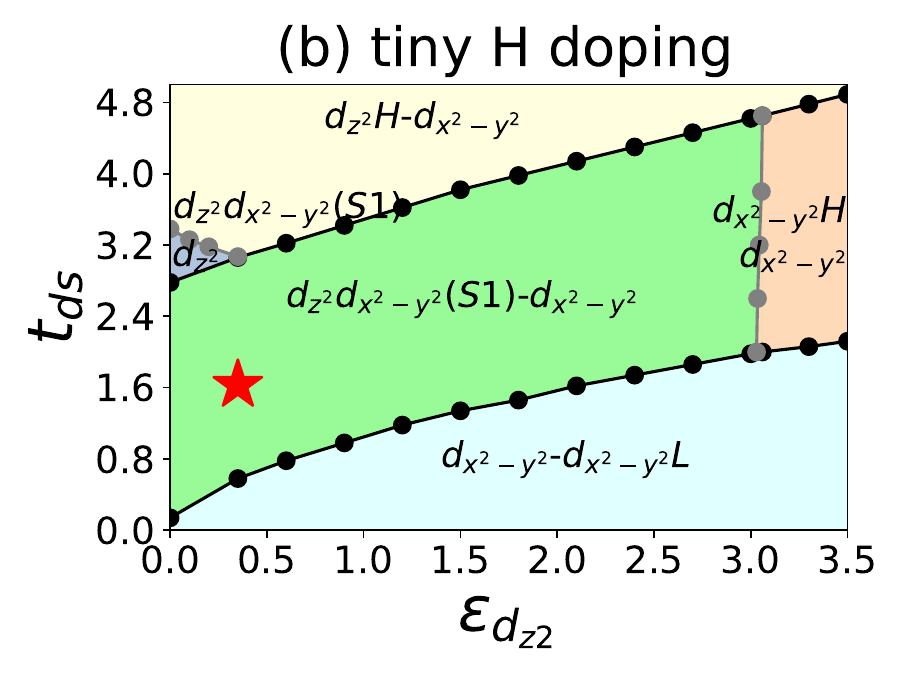,height=4cm,width=.33\textwidth, clip} 
\psfig{figure=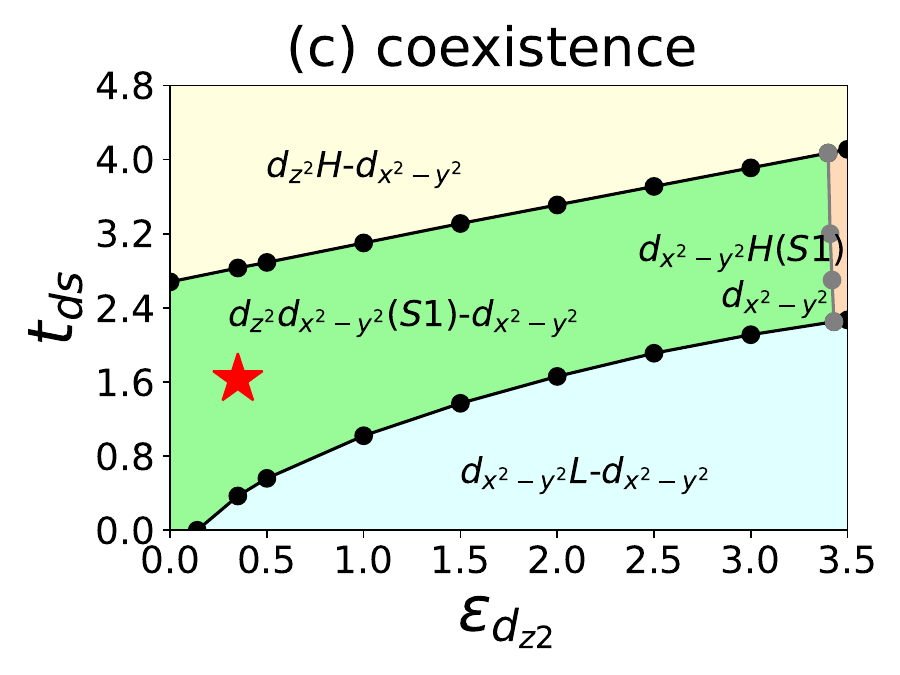,height=4cm,width=.32\textwidth, clip}

\caption{Phase diagram of undoped systems for (left) 100\% intercalated H; (middle) bilayer NiO$_2$ sandwiching a doped H models, and (right) coexistence of H doped and normal Ni with three holes totally. The red star denotes the location corresponding to realistic DFT parameter. $S1$ emphasizes the triplet $S=1$ nature of two-hole state.}
\label{phaseundoped}
\end{figure*}

In the absence of H doping, consistent with our previous studies~\cite{Mi2020}, the three models exhibit a common NiO$_4$ motif hosting a two-hole Zhang-Rice singlet (ZRS) \cite{zhang1988effective} at, for instance, $\epsilon_p$=3\,eV, which is typical for cuprates \cite{PhysRevB.93.155166}. This ZRS state is characterized by the dominant $d_{x^2-y^2}L$ weight. However, at sufficiently large $\epsilon_p$, the GS transits to a $d_{z^2}d_{x^2-y^2}$ triplet state ($S=1$) in accordance with Hund's rules \cite{PhysRevLett.126.127401}.
To investigate the influence of intercalated H doping, we first focus on the evolution of the GS weight distribution for the three models, as shown in Fig.\ref{weight} and Fig.\ref{phaseundoped}. These scenarios consider the situations where no additional hole doping is introduced apart from the self-doping effects on Ni and O orbitals caused by the intercalated H. Subsequently, we will analyze the hole-doped systems, as illustrated in Fig.\ref{weightdoped} and Fig.\ref{phasedoped}, and discuss the implications of our model calculations on the influence of H doping on superconductivity (SC). Note that in Fig.\ref{weight} and Fig.\ref{weightdoped}, the main figures display the total weight of general states, such as $d^8, d^9L$, etc., while the insets provide detailed weight distributions of dominant states, like $d_{z^2}d_{x^2-y^2}$, $d_{x^2-y^2}L$, etc.

\subsection{A.~Undoped systems with different concentration of topotactic H}

Figure~\ref{weight}(a) reveals that, for the DFT parameter $t_{ds}=1.63$ eV, which is relevant for infinite-layer nickelates, the GS with fully topotactic H is predominantly characterized by a $d_{z^2}d_{x^2-y^2}$ triplet ($S=1$) state (gray curve). This high-spin configuration is consistent with the electronic state observed in the topotactic H compound of LaNiO$_2$H, where a 3$d^8$ state was theoretically predicted~\cite{Si2019} and later observed experimentally~\cite{krieger2021charge,tam2021charge,raji2023charge}. Such a high-spin state typically favors a two-band Mott insulator under correlations~\cite{Si2019,Malyi2022}, which imposes significant constraints on the essential role of the Zhang-Rice-like $d^9L$ singlet in SC.
Moreover, apart from the dominant $d_{z^2}d_{x^2-y^2}$ triplet state, subdominant two-hole states like the triplet $d_{x^2-y^2}$H and $d_{z^2}$L states are unlikely to promote SC as well. Additionally, the singlet $d_{x^2-y^2}L$ state is only preserved when the value of $t_{ds}$ is unphysically small, i.e., $<$0.2 eV, suggesting that the hybridization between H-1$s$ and Ni-$d_{z^2}$ stabilizes the topotactic H by forming a bonding state between H-1$s$ and Ni-$d_{z^2}$.

In the case of tiny (single) H doping, mimicked by bilayer NiO$2$ with a single H as shown in Fig.~\ref{geom}(b), Fig.~\ref{weight}(b) reveals that the ground state of the H-induced self-doped system, with a total of three holes (as the single topotactic H absorbs one electron from the doubly $d^9$ configuration at two Ni sites), consists of a triplet $d_{z^2}d_{x^2-y^2}$ state in one layer and a $d_{x^2-y^2}$ state in the opposite layer. Importantly, this $d_{z^2}d_{x^2-y^2}$-$d_{x^2-y^2}$ state is degenerate with its parity $d_{x^2-y^2}$-$d_{z^2}d_{x^2-y^2}$ state, meaning their weights in the ground state are exactly the same.
This suggests that the system is stabilized by lowering its symmetry with inequivalent Ni impurities in opposite layers. In our model, where we emphasize the local physics using impurity treatment, the GS of $d_{z^2}d_{x^2-y^2}$-$d_{x^2-y^2}$ hints that the intercalated H only absorbs one electron from one of the two nearest Ni, simultaneously enhancing the on-site energy of Ni-$d_{z^2}$, resulting in $d_{z^2}d_{x^2-y^2}$ in one of the Ni sites. However, in realistic samples, this effect may be averaged due to partial occupancy of $d_{z^2}$ orbitals, forming the bonding state $d_{z^2}H$-$d_{z^2}$ and leading to a local two-band $e_g$ Ni dimer and an average state of $d^{8.5}$ with one hole at $d_{x^2-y^2}$ and half at $d_{z^2}$.
Hence, singly distributed H atoms without forming 1D H-chains, which simulates the case of low H doping (e.g.~$<$20\% \cite{ding2023critical}), may affect more neighboring Ni sites by promoting their non-singlet states and thereby disrupt the SC.

The above analysis of the two limiting models [Fig.~\ref{weight}(a,b)] demonstrates that both low and high concentrations of topotactic H are detrimental to SC in nickelates, aligning with the experimental observations in Ref.~\cite{ding2023critical}.
Now, we explore the most insightful situation of moderate H doping. Although the model we used contains only two Ni atoms, it effectively reflects the coexistence of H-chains surrounding Ni (3$d^8$) and nominal Ni (3$d^9$). The former represents the Ni atoms in-between H-chains; while the latter represents normal Ni atoms that are not in close proximity to H-chains.
Fig.~\ref{weight}(c) reveals that while the Ni impurity hybridized with the intercalated H always promotes the $d_{z^2}d_{x^2-y^2}$ triplet state, the separate Ni impurity can host a $d_{x^2-y^2}$ hole, which can move freely in the realistic lattice system and is essential for the phase-coherent SC. This critical observation is different from the previous two limiting models but qualitatively consistent with the experimental finding that moderate H doping supports the SC dome~\cite{ding2023critical}.
Another noteworthy feature in this two-Ni model is the smooth crossover between phases, which can be explained by the robust occupation at $d_{x^2-y^2}$ and one hole at $d^8$ Ni site gradually transferring from $L$ to $d_{z^2}$ as $t_{ds}$ increases.
These results indicate that the effects of H are very localized: it primarily affects the nearest Ni atoms, while the next-nearest Ni atoms are preserved as $d_{x^2-y^2}$ single occupancy, similar to the case of nominal Ni \cite{Jiang2019,PhysRevB.107.165116}.

\subsection{B.~Phase diagram of undoped systems with different concentration of topotactic H}

Figure~\ref{phaseundoped}(a-c) presents the phase diagram of the three models corresponding to Fig.~\ref{weight}(a-c), with the addition of the $\epsilon_{d_{z^2}}$ degree of freedom to provide supplemental insights into the different physics of these models, and how the topotactic H affects the GS. The red star in the phase diagram (a-c) denotes the relevant DFT parameters, which consistently fall within the associated phase region (green in Fig.~\ref{phaseundoped}) for all three characteristic concentrations of H doping.

Firstly, at around $\epsilon_{d_{z^2}}$=0.35\,eV as predicted by DFT and Wannier projections (Table~\ref{table}), all three models incorporate a component of the high-spin state $d_{z^2}d_{x^2-y^2}$ ($S=1$). For the slightly H doped case in Fig.~\ref{phaseundoped}(b) and the 50\% topotactic H case in Fig.~\ref{phaseundoped}(c), the other Ni sites are predominantly described by holes at $d_{x^2-y^2}$, indicating a coexistence of one- and two-band descriptions for the entire system. The robust occupation of the hole at $d_{x^2-y^2}$ in Fig.~\ref{phaseundoped}(b-c) suggests that the individual Ni atoms unhybridized with H are unaffected by the neighboring intercalated H.
In the case of small hybridization $t_{ds}$ between the (doped) H-1$s$ and $d_{z^2}$, the GS is characterized by the Zhang-Rice singlet (ZRS) \cite{zhang1988effective} with a dominant $d_{x^2-y^2}L$ component. However, when the $t_{ds}$ is strong enough [$>$3\,eV for $\epsilon_{d_{z^2}}$ in Fig.~\ref{phaseundoped}(b)], it results in the generation of $d_{z^2}$-$H$ bonding and anti-bonding states with a large energy separation, and the system prefers occupying the $d_{z^2}H$ state beyond a critical $t_{ds}$. 
On the other hand, for large enough $\epsilon_{d_{z^2}}$, the dominant component of the GS weight exhibits a crossover, shifting from $d_{z^2}$ to H orbitals due to the increasing energy cost of a large $\epsilon_{d_{z^2}}$.


\subsection{C.~Hole doped systems with different concentration of topotactic H}

\begin{figure*}[t!]
\psfig{figure=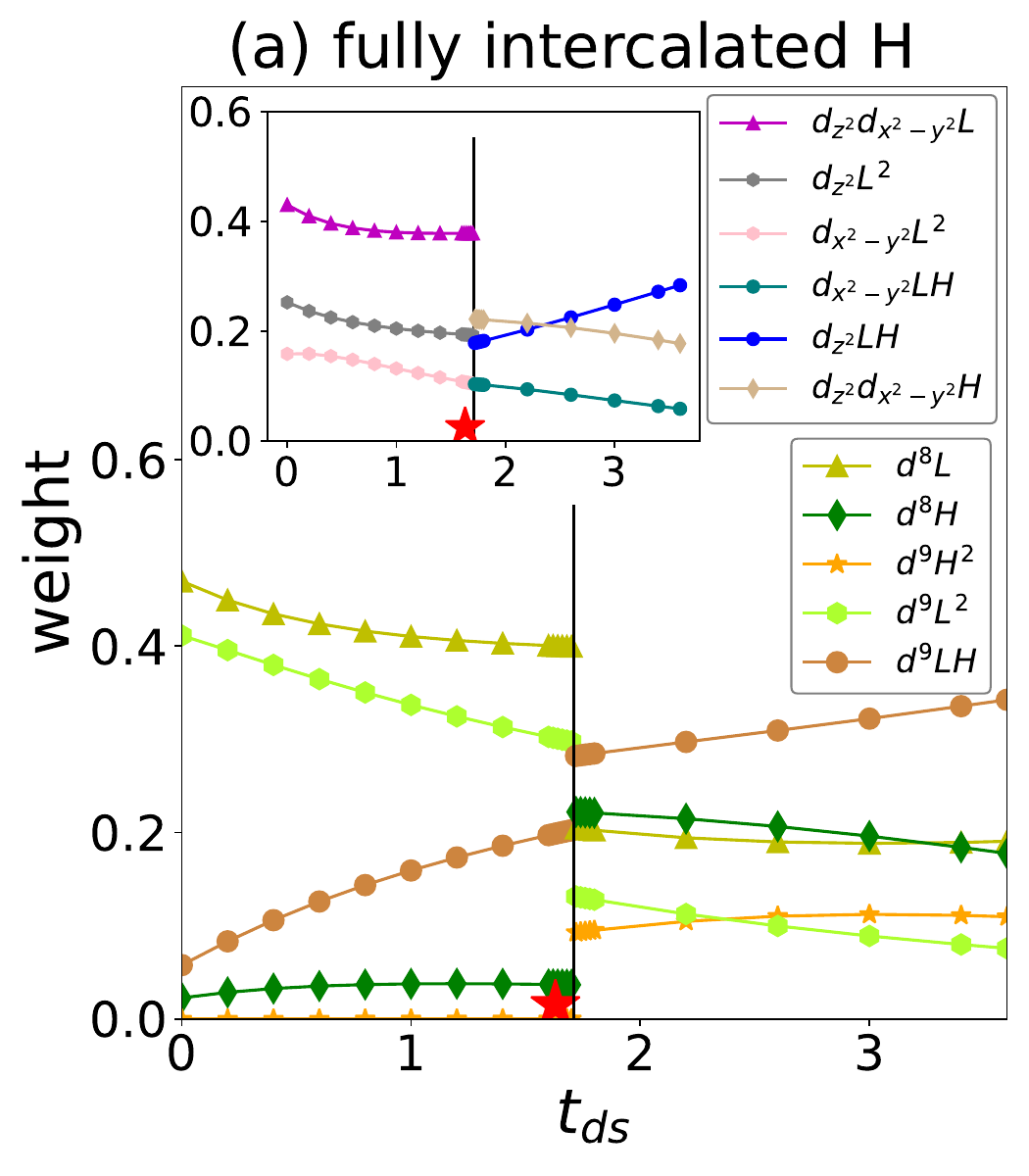,height=6cm,width=.31\textwidth, clip} 
\psfig{figure=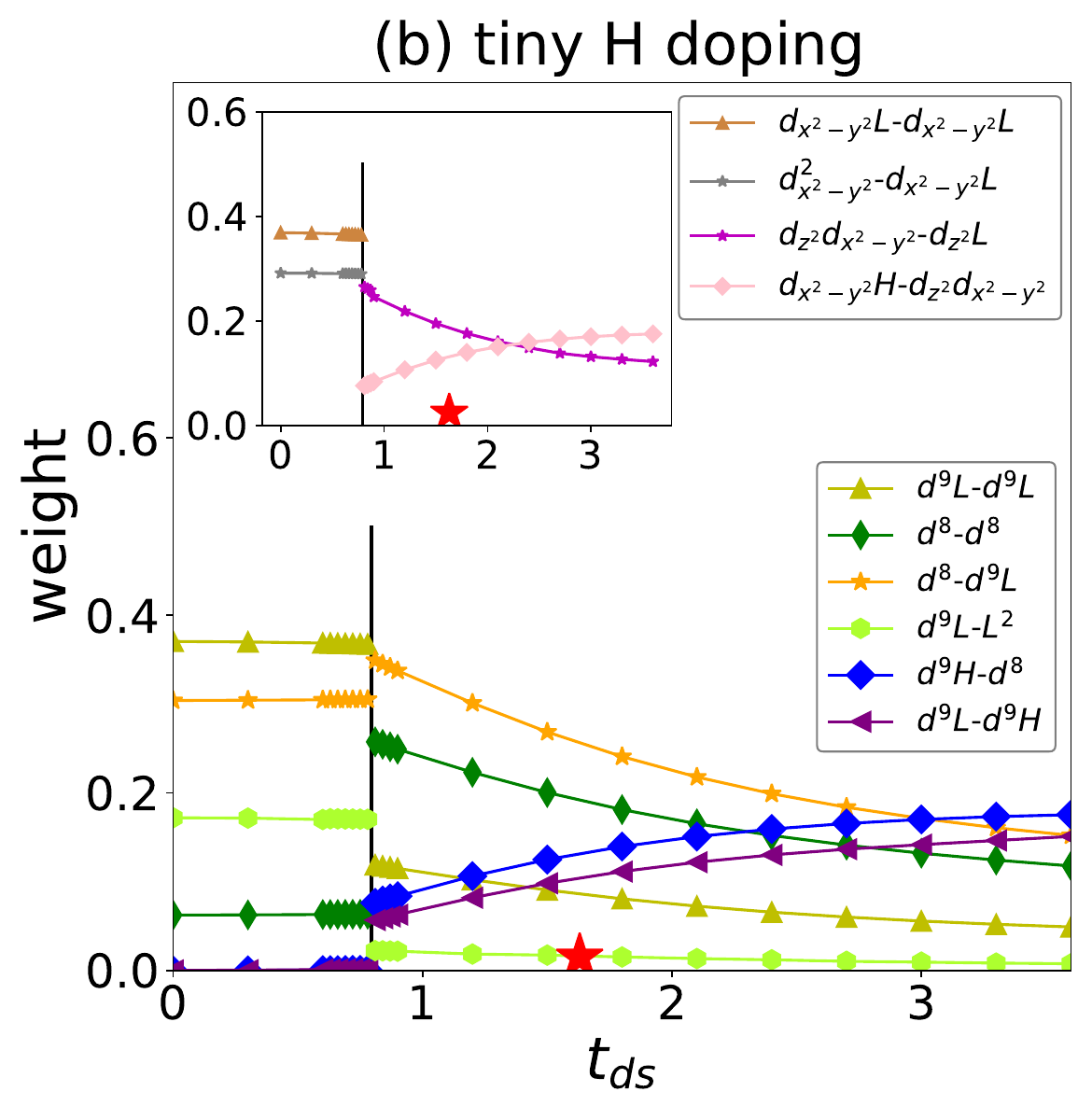,height=6cm,width=.31\textwidth, clip} 
\psfig{figure=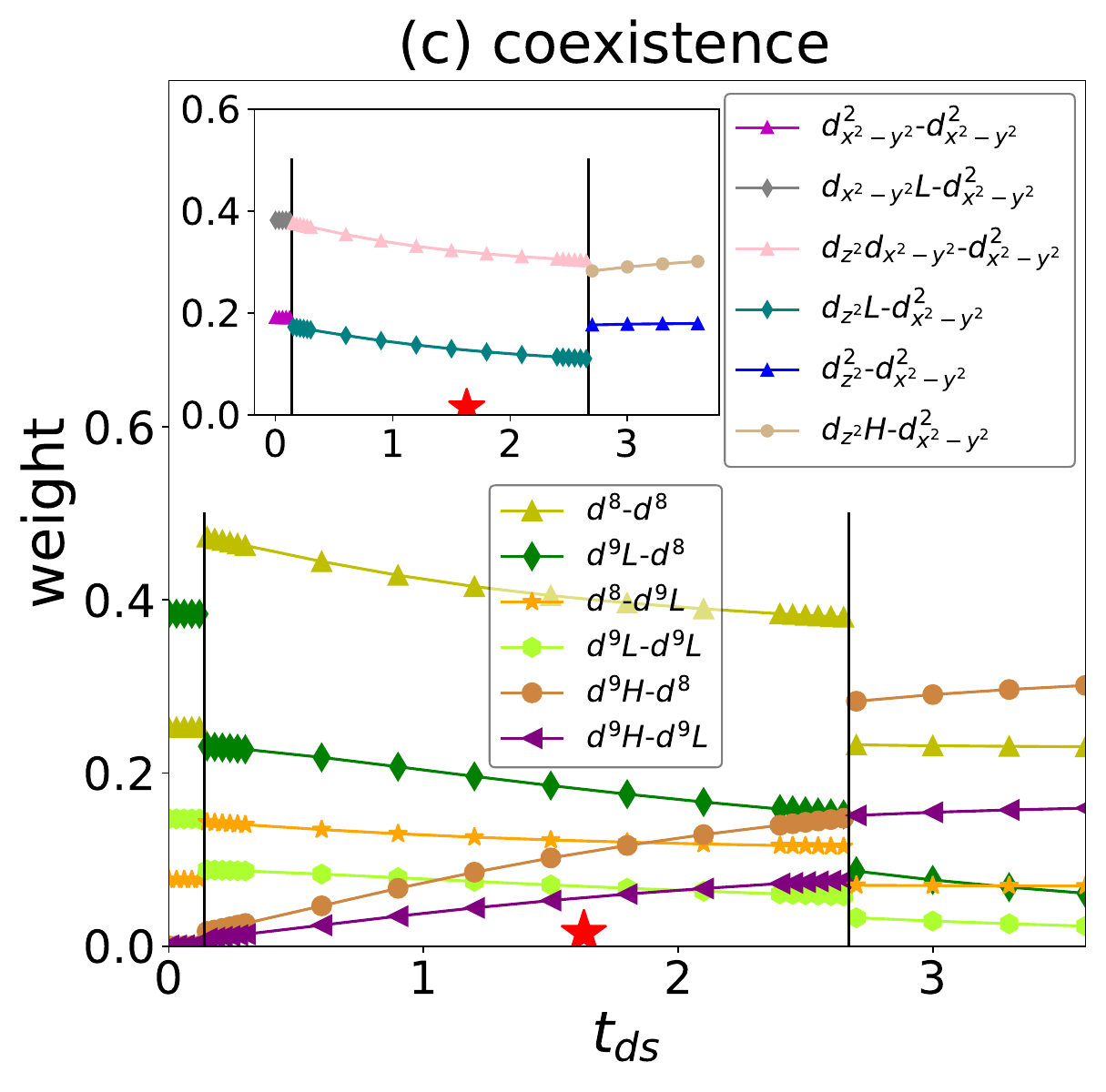,height=6cm,width=.31\textwidth, clip} 
\caption{Evolution of the GS weight distribution in doped H intercalated systems with increasing the hybridization $t_{ds}$ between Ni-$d_{z^2}$ and H-$s$ orbital akin to Fig.~\ref{weight}. The number of holes of all models are one more than that in Fig.~\ref{weight}.}
\label{weightdoped}
\end{figure*}

We discuss the effect of hole-doping, i.e., the addition of one more doped hole to the system, as shown in Fig.~\ref{weightdoped}. In the case of fully topotactic H situation depicted in Fig.~\ref{weightdoped}(a), when compared with Fig.~\ref{weight}(a), the additional doped hole resides either in the ligand $L$ or in the intercalated H. This additional hole does not modify the nature of the original two-hole $d_{z^2}d_{x^2-y^2}$ triplet. This observation strongly indicates that 100\% topotactic H would indeed strongly suppress SC by the persistence of the triplet component of $d_{z^2}d_{x^2-y^2}$.
Similarly, the comparison between Fig.~\ref{weight}(b) and Fig.~\ref{weightdoped}(b) reveals that, in the case of slight topotactic H doping, the single topotactic H located between the two Ni sites appears to break the inversion symmetry, leading to an asymmetric ground state in the two Ni sites. Specifically, one of the Ni sites becomes $d^8$, while the second becomes $d^9L$. Notably, the inset in Fig.~\ref{weightdoped}(b) further demonstrates that $d^8$ is predominantly composed of a $d_{z^2}d_{x^2-y^2}$ high-spin ($S=1$) triplet state, and $d^9L$ is achieved by a hole at $d_{z^2}$ and another hole at O site ($L$), forming a triplet state. Both the triplet natures of the two layers are therefore not conducive to SC. An interesting observation is that at stronger $t_{ds}$, the ground state becomes the $d_{x^2-y^2}H$-$d_{z^2}d_{x^2-y^2}$ state, where the H plays a similar role to the ligand O, due to the large energy separation between the $d_{z^2}$-H bonding and antibonding states.

The most intriguing case once again occurs for the moderate topotactic H situation, as depicted in Fig.~\ref{weightdoped}(c), where the $d^8$-$d^8$ states dominate the DFT parameter region. Now, the doped hole resides at the nominal Ni$^{1+}$ site, forming a $d^2_{x^2-y^2} \equiv d_{x^2-y^2}d_{x^2-y^2}$ singlet, which is crucial for the phase-coherent SC, demonstrating that in topotactic H nickelates the doped holes are prone to occupy the $d_{x^2-y^2}$ of Ni$^{1+}$. Similar $d^8$ singlet state had been observed in Ref.~\cite{PhysRevB.104.L220505}. This indicates that when there is a moderate concentration of topotactic H and 1D H-chains are formed, the doped hole induces inter-site hopping between the nominal Ni$^{1+}$ and the high-spin Ni$^{2+}$. In Ni$^{2+}$, the $d_{z^2}$ orbital is expected to receive the hole from Ni$^{1+}$'s $d_{x^2-y^2}$ orbital because the energy of the $d_{z^2}$ orbital (in electron language) is lower than that of the $d_{x^2-y^2}$ orbital and such hopping is unfavorable to nickelate SC.

\subsection{D.~Phase diagram of hole doped systems with different concentration of topotactic H}

\begin{figure*}[t!]
\psfig{figure=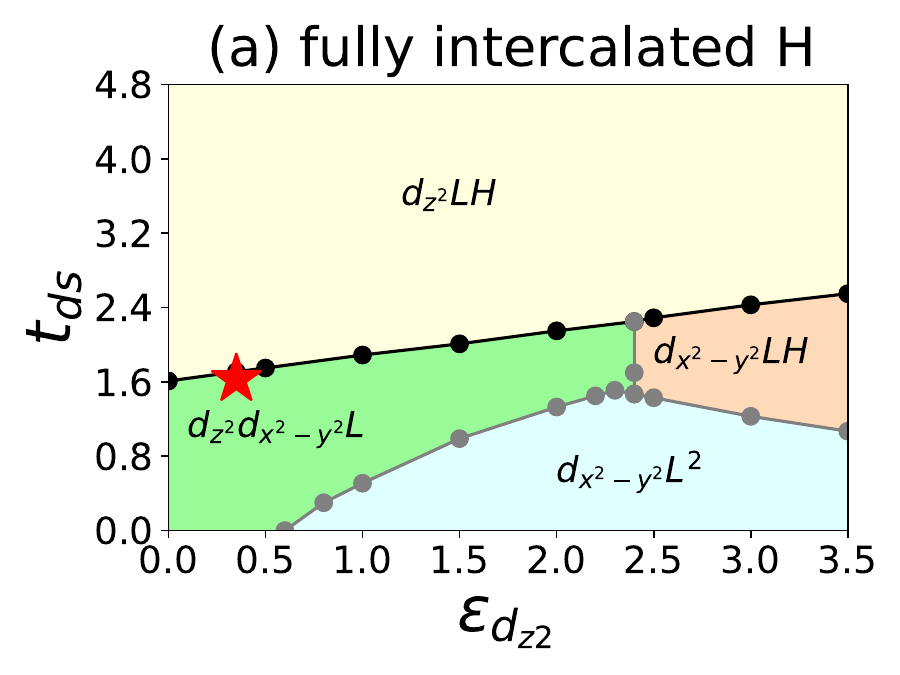,height=4cm,width=.32\textwidth, clip}
\psfig{figure=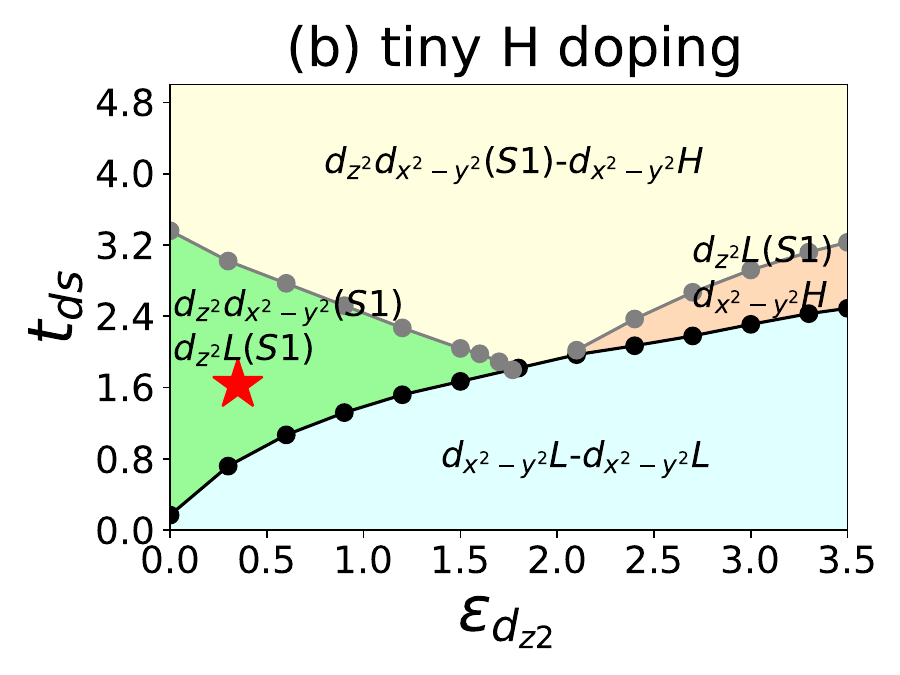,height=4cm,width=.33\textwidth, clip} 
\psfig{figure=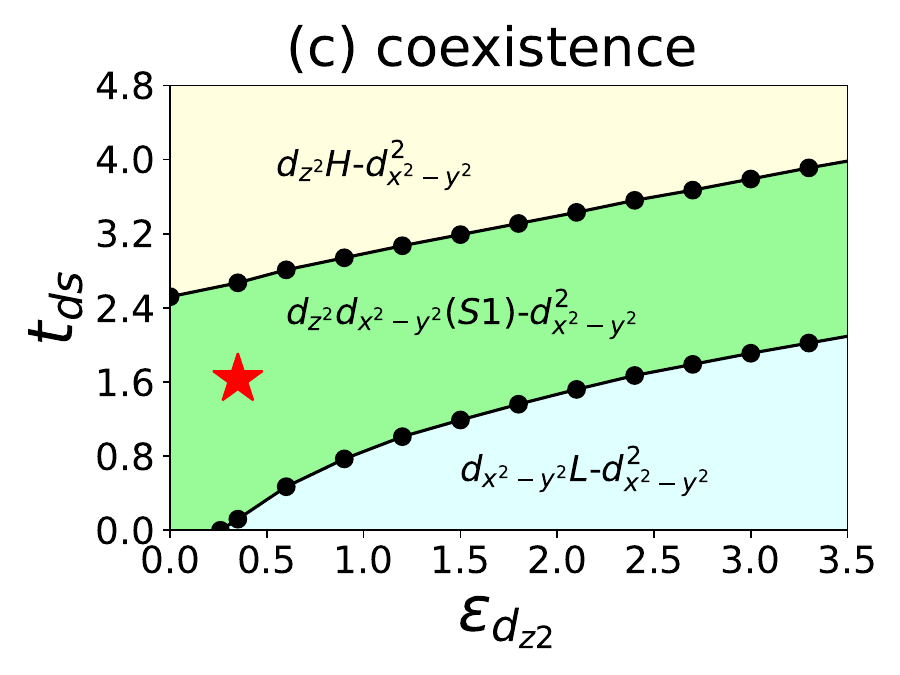,height=4cm,width=.32\textwidth, clip} 

\caption{Phase diagram of doped systems for three models with four holes totally akin to Fig.~\ref{phaseundoped}. $S1$ emphasizes the triplet $S$=1 nature of two-hole state at $d_{x^2-y^2}$ and $d_{z^2}$.}
\label{phasedoped}
\end{figure*}

Figure~\ref{phasedoped}(a-c) presents the phase diagram of the three models corresponding to Fig.~\ref{weightdoped}(a-c).
In the case of the fully topotactic H situation [Fig.~\ref{phasedoped}(a)], despite the DFT-relevant $\epsilon_{d_{z^2}}$=0.35\,eV being located near the phase boundary, the doped hole always resides in the ligand $L$ regardless of the magnitude of $t_{ds}$, as compared with Fig.~\ref{phaseundoped}(a). When the $\epsilon_{d_{z^2}}$ goes from small to large values, the hole transits from $d_{z^2}$ to H-1$s$ ($H$).
This observation indicates that additionally doped hole in fully topotactic H system occupies either the $d_{z2}$ or H-1$s$, depending on the potential of $d_{z^2}$, however, both GS compete with SC, making it unfavorable.

In another two models shown in Fig.~\ref{phasedoped}(b,c), the red stars are within the specific phase regimes, ensuring that the GS illustrated in Fig.~\ref{weightdoped} does not sensitively depend on the detailed parameter choices. Similar to Fig.~\ref{phaseundoped}(b,c), at small hybridization $t_{ds}$ between the doped H and $d_{z^2}$, the GS is characterized by the Zhang-Rice singlet (ZRS) \cite{zhang1988effective} with a dominant $d_{x^2-y^2}L$ or $d^2_{x^2-y^2}$ component. Conversely, when $t_{ds}$ is strong enough, it tends to locate the hole onto H by forming $d_{x^2-y^2}H$ and $d_{z^2}H$ in Fig.~\ref{phasedoped}(b) and (c), respectively.

On the other hand, for large enough $\epsilon_{d_{z^2}}$, the dominant component of the GS weight undergoes a crossover, transferring from $d_{z^2}$ to H orbitals due to the increasing energy cost of a large $\epsilon_{d_{z^2}}$, except in the case of 50\% H doping, where the green regime does not host any dominant states with holes on intercalated H, the additional hole goes to the $d_{x^2-y^2}$ at nominal N$^{1+}$.

\subsection{E.~Implication for H doped superconductivity}

\begin{figure*}[t!]
\begin{minipage}{.795\textwidth}
\centering
\includegraphics[width=0.95\textwidth]{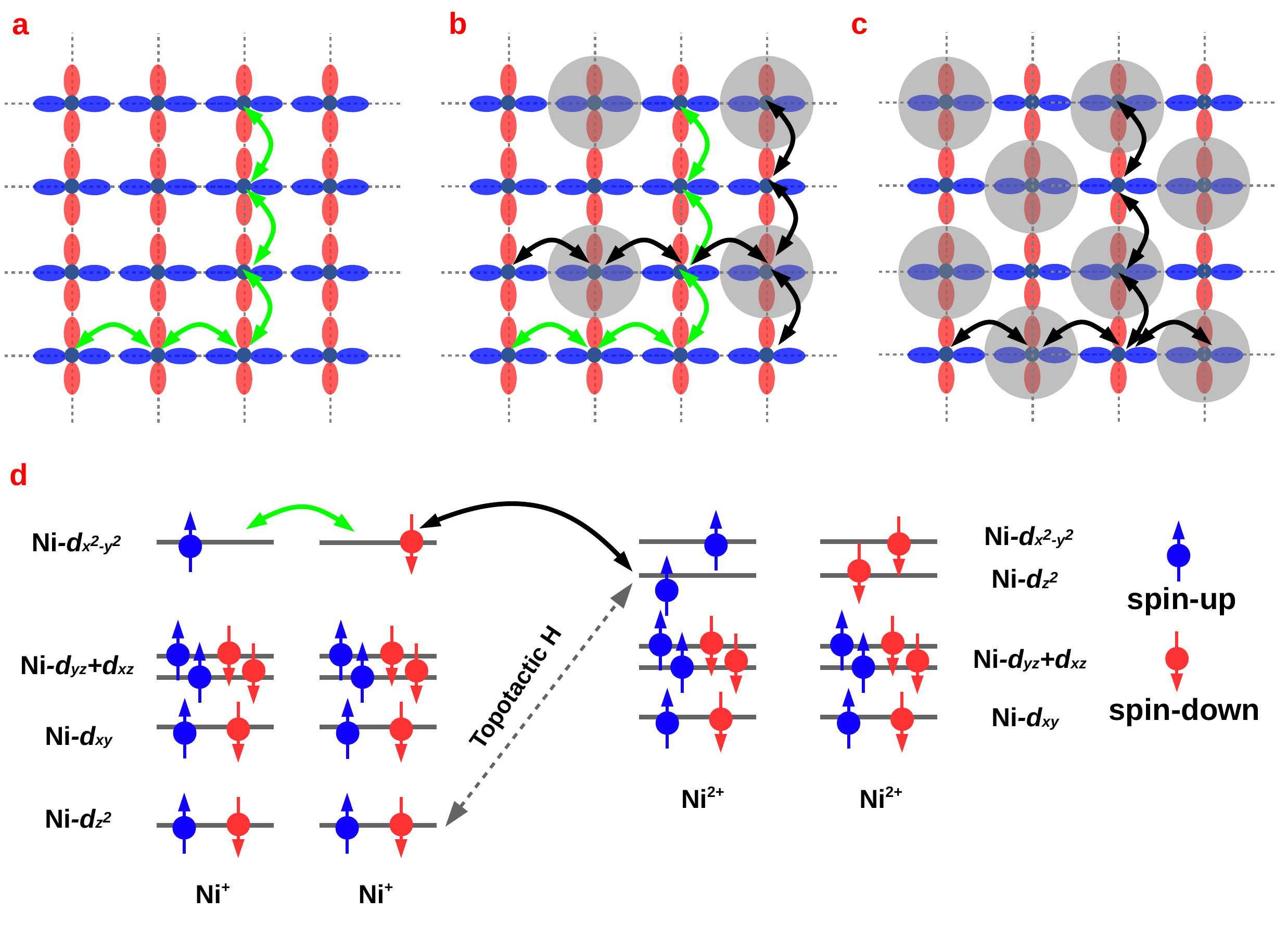}
\end{minipage}\hfill
\begin{minipage}{.195\textwidth}
\caption{Schematic diagrams of how the nickelate SC is preserved in samples with (a) 0\% and (b) 25\%, and (c) destroyed in 50\% topotactic H. The green and black arrows indicate effective inter-site hopping between pure singlet state of $d_{x^2-y^2}$, and between singlet and $d_{z^2}d_{x^2-y^2}$ triplet ($S$=1) state (gray circles), respectively. (d) Schematic figure showing the electron hopping process in Ni$^+$ and Ni$^{2+}$ states.}
\label{fig6}
\end{minipage}
\end{figure*}

Based on all the previous numerical evidence, we can now provide deeper physical insights into the impact of H doping on the SC of infinite-layer nickelates.

When the parent compound, such as LaNiO$_2$ or NdNiO$_2$, is away from topotactic (re-intercalation) H, the system is physically described by a single Hubbard band of $d_{x^2-y^2}$, as shown in Fig.~\ref{fig6}(a) \cite{Held2022}. However, for topotactic H systems, the energy level of $d_{z^2}$ is enhanced by topotactic H, leading to the reduction of Ni$^{1+}$ to Ni$^{2+}$ because H behaves as H$^-$, absorbing one electron \cite{Si2019}.

In the optimal H doping regime, i.e., when the concentration of topotactic H approaches 25\% \cite{ding2023critical}, the formation of 1D H-chains perpendicular to the Ni-O planes allows for the coexistence of $d_{x^2-y^2}$ of nominal Ni$^{1+}$, and a high-spin ($S=1$) $d_{z^2}d_{x^2-y^2}$ triplet state, as shown in Fig.~\ref{fig6}(b). The further hole doped by Sr or Ca substitution on nominal Ni$^{1+}$ is prone to occupy $d_{x^2-y^2}$, leading to $d^{9-\delta}$ states ($\delta$$\sim$0.2 for optimal SC). This coexistence results in two different types of inter-site hopping processes: the hoppings [green arrows in Fig.~\ref{fig6}] between Ni$^{1+}$ sites, which construct long-range SC, and the hoppings (black arrows in Fig.~\ref{fig6}) from Ni$^{1+}$ $(d_{x^2-y^2})$ to Ni$^{2+}$ sites [highlighted by gray in Fig.~\ref{fig6}(b,c)], which are unfavorable for SC since $d_{z^2}$ within Ni$^{2+}$ is energetically lower than $d_{x^2-y^2}$. In Fig.~\ref{fig6}(c), when the concentration of topotactic H exceeds 25\%, for instance, approaching 50\%, only the inter-site hoppings from $d_{x^2-y^2}$ to $d_{z^2}$ dominate the system, and the SC is completely destroyed.

Fig.~\ref{fig6}(d) provides supplemental and schematic energy diagrams, illustrating the enhancement of the on-site energy ($\epsilon_{d_{z^2}}$) of the $d_{z^2}$ orbital due to the presence of topotactic H and the consequent hybridization $t_{ds}$, which leads to distinct inter-site electron hopping processes in Ni$^{1+}$ and Ni$^{2+}$ states. Therefore, the three characteristic models designed in this study provide a generic physical picture of the influence of topotactic H on superconductivity.

\section{IV.~Summary and outlook}

Motivated by the recent experimental discovery \cite{ding2023critical} that the hydrogen incorporation appears critical for the infinite-layer nickelate SC, we performed a theoretical combination of DFT and impurity approximation to explore three designed characteristic multi-orbital Hubbard models to represent the high, low and moderate concentrations of topotactic hydrogen respectively. Our impurity models treat Ni and H as impurities embedded in the O lattice and properly incorporate the local $3d^8$ multiplet structure of Ni impurity with full Coulomb and exchange interactions. The effects induced by topotactic H can be classified to the following aspects:

(1) Our simulations indicate that high concentrations of topotactic hydrogen induce pure high-spin states ($S$=1) that is composed by $d_{z^2}d_{x^2-y^2}$, which is consistent with the experimental observation in Ref.~\cite{tam2021charge,krieger2021charge,raji2023charge}. The Mott insulating nature of the hydroxide makes the electrons localized and hence the hopping between $d_{x^2-y^2}$ is eliminated. Therefore, such a high concentration of topotactic H is unfavorable to SC in nickelate. Additionally doped holes are expected to be trapped by O-2$p$ ($L$) or H-1$s$ ($H$), further suppressing the nickelate SC. 

(2) For low concentration of H, which is simulated by intercalating single H into a bilayer NiO$_2$ model, our simulations demonstrate that the topotactic H breaks the bilayer inversion symmetry by resulting in $d^8$ ($d_{z^2}d_{x^2-y^2}$)  and $d^9$ ($d_{x^2-y^2}$) states in each layer, respectively. In realistic samples, such a symmetry breaking may be averaged by temperature or lattice vibration, leading to $d^{8.5}$ ($d_{x^2-y^2}d^{0.5}_{z^2}$) in both site. Another possibility is the two electronically different states coexist and local crystal disorder or atomic defect would possibly play a role at pining one state in one Ni. This symmetry breaking induced by low concentration of topotactic H may play as driving force of the experimentally observed charge order state \cite{tam2021charge,krieger2021charge,raji2023charge} and the absence of magnon in non-capped nickelates \cite{krieger2021charge}.

(3) The optimal concentration of 25\% H matches with the single Ni-$d_{x^2-y^2}$ band picture of SC in infinite-layer nickelates. Specifically, the out-of-plane ordered formation and in-plane distribution of one-dimensional H-chains \cite{PhysRevB.107.165116} perpendicular to Ni-O plane protects the one-hole $d_{x^2-y^2}$ undoped state and further two-hole $d^2_{x^2-y^2}$ singlet states at nominal Ni$^{1+}$ by limiting the impact of topotactic-H to Ni sites in between H. Hence, the long-range effective inter-site electron/hole hopping between $d_{x^2-y^2}$ is partially allowed along both $x$ and $y$ direction [Fig.~\ref{fig6}(b)] and thereby beneficial to SC upon additionally/slightly hole doping at H$\sim$25\%. On the contrary, when the concentration of H exceeds $\sim$50\%, the additionally doped hole is still trapped by the nominal Ni$^{1+}$ at $d_{x^2-y^2}$ orbital, leading to the formation of $d^2_{x^2-y^2}$ state, whose existence would seemingly protect the SC. Nonetheless, due to the large amount of high-spin Ni$^{2+}$, the effective inter-site hopping between $d_{x^2-y^2}$ orbitals in the hole doped system is totally cut off along both $x$ and $y$ directions [Fig.~\ref{fig6}(c)] while only those inter-site hoppings between $d_{x^2-y^2}$ to $d_{z^2}$, that are detrimental to SC, are expected. Therefore, both undoped and doped cases at H$\sim$50\% are not beneficial to SC. Additionally, the existence of inter-site hopping between $d_{x^2-y^2}$ to $d_{z^2}$, that are unfavorable to SC, may explain why $T_c$ in nickelate SC are more sensitively dependent upon the sample. In fact, some of the superconducting samples were discovered to host non-zero resistivity even when $T$$\sim$0\,K \cite{wang2022pressure}. Besides, it might also be closely related to the observation that the transition temperature range is generally broader than that in other superconductors \cite{li2019superconductivity,Li2020,zeng2020}.

\section{ACKNOWLEDGMENTS}
We would like to thank Karsten Held for illuminating discussions.
C.~Q.~and M.~J.~acknowledge the support by National Natural Science Foundation of China (NSFC) Grant No.~12174278, startup fund from Soochow University, and Priority Academic Program Development (PAPD) of Jiangsu Higher Education Institutions. 
L.~S.~is thankful for the starting funds from Northwest University.
L.~S.~also acknowledges the financial support by projects P32044 and I5398 of the Austrian Science Funds (FWF).
Calculations have been done mainly on the Soochow University and Super-computing clusters at Northwest University.

\bibliography{main}

\end{document}